\documentclass[english]{article}
\usepackage[T1]{fontenc}
\usepackage[latin9]{inputenc}
\usepackage{amsmath}
\usepackage{cite}
\usepackage[normalem]{ulem}

\usepackage{color}

\makeatletter
\usepackage{amssymb}
\usepackage{latexsym}
\usepackage{epsfig}
\usepackage{float}
\usepackage{dsfont}

\newcommand{\be}{\begin{equation}}
\newcommand{\ee}{\end{equation}}

\allowdisplaybreaks

\makeatother

\usepackage{babel}
\begin{document}
{}~ \hfill\vbox{\hbox{CTP-SCU/2017011}}\break
\vskip 3.0cm
\centerline{\Large \bf  Fixing the AdS$_3$ metric from  the pure state entanglement entropies of CFT$_2$}

\vspace*{10.0ex}
\centerline{\large Peng Wang, Houwen Wu and Haitang Yang}
\vspace*{7.0ex}
\vspace*{4.0ex}
\centerline{\large \it College of Physics}
\centerline{\large \it Sichuan University}
\centerline{\large \it Chengdu, 610064, China} \vspace*{1.0ex}
\vspace*{4.0ex}

\centerline{pengw@scu.edu.cn, iverwu@scu.edu.cn, hyanga@scu.edu.cn}
\vspace*{10.0ex}
\centerline{\bf Abstract} \bigskip \smallskip
In this paper, based on RT formula, by identifying the pure state UV and IR entanglement entropies of a perturbed CFT$_2$ with geodesic lengths in the bulk, we demonstrate that the dual geometry is uniquely determined to be asymptotically AdS$_3$. 
The pure AdS$_3$ geometry is recovered by taking the massless limit of the system. Our derivations hold in both static and covariant scenarios.

\vfill
\eject
\baselineskip=16pt
\vspace*{10.0ex}

\section{Introduction}
The AdS/CFT correspondence plays a central role in modern theoretical physics. This conjecture states that a weakly coupled gravitational theory in a $d+1$ dimensional Anti-de Sitter (AdS) space is equivalent to a strongly coupled $d$ dimensional conformal field theory (CFT) on its conformally flat boundary \cite{Maldacena:1997re}. It provides a testable realization of the holographic principle \cite{tHooft:1993dmi,Susskind:1994vu}. 

Quantum entanglement is a manifestation of the non-local property of quantum mechanics. 
For the simplest configuration, a quantum system is divided into two parts: $A$ and $B$. The Hilbert space is thus decomposed into $\mathcal{H}=\mathcal{H}_{A}\otimes\mathcal{H}_{B}$. The entanglement entropy of region $A$ is defined as the von Neumann entropy, $S_{A}=-\mathrm{Tr}\left(\rho_{A}\ln\rho_{A}\right)$, where $\rho_{A}$ is the reduced density matrix of region $A$: $\rho_{A}=\mathrm{Tr}_{\mathcal{H}_{B}}\rho$. It is evident that $S_{A}=S_{B}$ for this pure state. One of the most successful supports of the AdS/CFT correspondence is the Ryu-Takayanagi (RT) formula, which asserts the equality of the entanglement entropy (EE) of the boundary CFT and the accordingly defined minimal surface area in the bulk AdS \cite{Ryu:2006bv,Ryu:2006ef,Hubeny:2007xt}. In their work, the geodesic length (minimal surface area) in AdS$_{3}$ is calculated and found to agree with the EE of CFT$_{2}$. This identification has been extensively verified by subsequent studies, as summarized in a review \cite{Rangamani:2016dms} and references therein. 

Motivated by the success of the RT formula, a conjecture has been proposed suggesting that gravity can be interpreted as an emergent structure, determined by the quantum entanglement of the dual conformal field theory (CFT) \cite{VanRaamsdonk:2009ar,VanRaamsdonk:2010pw}. This idea was further developed by Maldacena and Susskind, who proposed an equivalence between the Einstein-Rosen bridge (ER) and the Einstein-Podolsky-Rosen (EPR) experiment \cite{Maldacena:2013xja}. Building on this framework, a crucial question arises: \emph{Can we determine the leading behavior of the dual bulk geometry, specifically the metric, from the entanglement entropies of the CFT?} 

This question has attracted considerable attention in recent years. The primary challenge lies in determining the metric of the holographic dimension that extends into the bulk. It is widely accepted that this hidden extra dimension is generated by the energy cut-off of the CFT and oriented perpendicular to the boundary, as realized by the holographic principle \cite{Susskind:1998dq}. To date, two major approaches have been developed to tackle this problem: The first
one is the tensor network which generates a logarithmic geodesic length from boundary states, leading to a discretized AdS space \cite{Swingle:2009bg}. These networks provide an intuitive connection between entanglement and spatial geometry.
The second method resorts to \emph{kinematic space}. This method maps geodesics in AdS$_3$ to points in a kinematic space. Specifically, the Crofton form, defined as the second derivative of entanglement entropy with respect to two distinct points (without taking the coincidence limit), plays a crucial role. A codimension-1 locus, or "point curve," in kinematic space corresponds to a family of geodesics that intersect at a single point in AdS$_3$. The geodesic length between any two points in the bulk is then obtained by integrating the Crofton form over the area between the corresponding point curves in the kinematic space \cite{Czech:2015qta}. The advantage of this approach lies in its clear correspondence between Crofton's formula and field theory, making it possible to derive operator product expansion (OPE) blocks from local operators in AdS via the Radon transformation \cite{Czech:2016xec,deBoer:2016pqk}. However, despite its elegance, this method is not straightforward when it comes to fixing the leading behavior of the spacetime metric in an unknown bulk geometry. In contrast to these approaches, our method introduces a key difference: we impose the limit $x \to x'$ after taking the derivatives. This procedure allows us to directly obtain the metric from the geodesics, offering a more immediate pathway to deducing the bulk geometry. In addition to these advancements, significant progress has also been made in the dynamics of the system, such as deriving linearized Einstein equations from entanglement entropy \cite{Faulkner:2013ica}.

In this paper, 
we develop a method to explicitly determine the leading behavior of the dual geometry from the free $\mathrm{CFT_2}$. At first glance, one might consider this problem trivial, citing symmetry arguments since both $\mathrm{CFT_2}$ and AdS$_3$ share the same $SO(2,2)$ symmetry. However, 
as is commonly known, all $d$-dimensional CFTs share the same $SO(2,d)$ symmetry, which their dual geometries must also respect.
So, symmetry matching is a necessary condition, but not a sufficient one.
Therefore, it becomes crucial to develop methods that go beyond symmetry arguments and provide a more robust framework. The approach we propose in this work satisfies several key requirements: The first requirement of the expected new method is to reproduce pure AdS$_{d+1}$ uniquely
from the free $\mathrm{CFT_d}$, ensuring a solid foundation for the correspondence. 
The second step is to reconstruct  nontrivial topologies
from the dual CFT. 
The next step is to build  one-to-one correspondences
between the excited states of the CFT and their associated bulk geometries.  
The final step is to derive the dynamics,
namely the Einstein equation, directly from the CFT data. These steps form a systematic pathway toward not only confirming the duality but also deepening our understanding of the holographic principle and its applications in emergent spacetime physics.

In contrast to previous efforts in the literature, our strategy is based on two key distinctions: Synge's world function \cite{Synge:1960} and the infrared entanglement entropy (IR EE)\footnote{In this paper, we refer to the entanglement entropy of a perturbed (mass-gapped) CFT$_2$ as IR EE for simplicity, distinguishing it from the traditional ultraviolet entanglement entropy (UV EE).}. Constructing the geodesic length is inherently simpler than directly deriving the metric, a point acknowledged in earlier work on integral geometry and kinematic space \cite{Czech:2015qta}. However, the significance of Synge's world function appears to have been overlooked in those approaches. While previous authors utilized derivatives with respect to two distinct points along geodesics, they did not apply the coincidence limit to extract the metric. In classical gravity, Synge's world function, defined as one half of the squared geodesic length, is a fundamental tool for studying the motion of a self-drived particle in a curved background. Almost all significant two-point quantities in classical gravity are derived from this function. For our current problem, the critical importance of Synge's world function lies in its ability to compute the metric directly. By taking second derivatives of the function and applying the coincidence limit, we can bypass irrelevant complexities and obtain the desired results with precision. This approach eliminates the ambiguities often associated with reconstructing the bulk geometry. Additionally, the parallels between the kinematic space formalism and Synge's world function are noteworthy. Although the two frameworks have distinct focuses, their similarities could inspire further developments.

The second critical quantity we incorporate is the IR entanglement entropy (IR EE) of a perturbed CFT$_{2}$. Traditionally, attempts to construct the dual geometry have relied only on the UV EE but have not been successful. Intuitively, reducing from a higher-dimensional system to a lower-dimensional one is straightforward, whereas reconstructing higher-dimensional information from lower-dimensional data is inherently more challenging. On the surface, the UV EE is fully expressed in terms of flat CFT quantities, with no apparent signature of the hidden bulk dimension. This suggests that the UV EE alone is insufficient for reconstructing the dual geometry. One possible explanation is that the UV EE inherently depends on a UV cut-off, intrinsic to quantum field theories. However, on the (bulk) classical gravity side, this UV cut-off can, in principle, be removed. For a massless CFT, although there is no intrinsic IR cut-off, the correlation length diverges, rendering the IR EE not directly expressible. This observation leads us to conjecture that introducing a finite IR cut-off could be of help. By progressively pushing this cut-off to infinity, it is possible to extract meaningful results without introducing inconsistencies on the gravity side. This approach is similar to the Pauli-Villars regularization in quantum field theory, where a large cut-off is introduced to isolate finite physical results. For a perturbed CFT with a finite IR cut-off, we anticipate that the leading-order behavior of the dual geometry corresponds to an asymptotic AdS space. The finite IR cut-off provides a natural framework for encoding the bulk geometry, bridging the gap between the perturbed CFT and its holographic dual.

Our derivations   demonstrate that the frequently used UV entanglement entropy (EE) of the $\mathrm{CFT_2}$ can only provide partial information about the asymptotic behavior of the boundary directions. However, it does not offer any direct insight into the energy-generated direction. Remarkably, we find that the entanglement entropy in the IR region of a perturbed CFT$_2$ provides the sufficient condition to determine both the energy-generated direction and the residual freedom in the boundary directions. This is because the IR EE is determined by both the UV and IR energy scales, which correspond to different values along the energy-generated direction. Therefore, by combining both the UV and IR EEs, we can uniquely determine the asymptotic form of the geodesic length in the bulk. This, in turn, leads to the determination of the metric of the dual geometry, which is found to be the anticipated asymptotic AdS$_3$. The pure AdS$_3$, which serves as the gravity dual of the massless CFT$_2$, is obtained by taking the mass scale of the perturbed CFT$_2$ to the massless limit. This corresponds to taking the correlation length of the perturbed CFT$_2$ to infinity. Moreover, the covariant case can be derived using the same approach, following the same underlying pattern.


\section{Spacetime metric from entanglement entropy}
Before starting the discussion,  we summarize the methods and steps.

\vspace{2em}

\subsection*{Strategy:}

\noindent \textbf{Main Equation: Synge's World Function}, Synge's world function is instrumental in extracting the spacetime metric from geodesics. It is crucial to note that the geodesics referred  here are those in the bulk geometry, while the known entanglement entropies correspond to boundary-attached geodesics. Therefore, our strategy involves deducing the general expression for bulk geodesics based on boundary-attached geodesics and then employing Synge's world function to obtain the metric. Our steps are given as follows:

\begin{enumerate}
\item  Use the Ryu-Takayanagi (RT) formula to determine two types of boundary-attached geodesics from the entanglement entropies.
\item Write the general expression for the bulk geodesics.
\item Fix the free coefficients and functions of the bulk geodesics using the known boundary-attached geodesics.
\item  Use Synge's world function to compute the spacetime metric.
\end{enumerate}

\vspace{2em}

\subsection*{Synge's world function:}

In classical gravity, the Synge's world function plays a fundamental role in investigating the radiation back-reaction (self-force) of a particle moving in a curved background. All the bi-tensors are defined by the Synge's world function. A comprehensive review on this subject can be found in \cite{Poisson:2011nh}. We only list some useful results here. Given a fixed point on a manifold $M$, and another point $x^{\prime}$ which connects to $x$ through a single geodesic $x=x\left(\tau\right)$, $\tau\in\left[0,t\right]$, such that  $x\left(0\right)=x$ and $x\left(t\right)=x^{\prime}$, the Synge's world function is defined as the square of the geodesic length:

\begin{equation}
\sigma\left(x,x^{\prime}\right)=\frac{1}{2}L_{\gamma_{A}}^{2}\left(x,x^{\prime}\right)=\frac{1}{2}t\int_{0}^{t}d\tau g_{ij}\dot{x}^{i}\dot{x}^{j},
\end{equation}

\noindent where $L_{\gamma_{A}}$ is the geodesic length connecting the points $x$ and $x^{\prime}$. This function is a bi-scalar for the points $x$ and $x'$, respectively. Throughout this paper, we use $i$ and
$i^{\prime}$ to distinguish   two points $x$ and $x'$ . The first derivative with respect to $x$ or $x'$ is the ordinary derivative, denoted as $\sigma_{i}=\frac{\partial\sigma}{\partial x^{i}}$ or
$\sigma_{i^{\prime}}=\frac{\partial\sigma}{\partial x^{i^{\prime}}}$.
It should be noted that $\sigma_{i}(x,x')$ $(\sigma_{i^{\prime}})$
is a vector for point $x$ $(x')$ but a scalar for point $x'$ $(x)$.
It is not   hard to find

\begin{equation}
\sigma_{i}(x,x')=t\,\dot{\frac{d}{dt}x_{i},\qquad\sigma_{i'}(x,x')=-t\,\frac{d}{dt}x_{i'}.}
\end{equation}

\noindent As usual, the second derivative at a single point is interpreted as the covariant derivative: $\sigma_{ij}\equiv\nabla_{j}\sigma_{i}$ and $\sigma_{i^{\prime}j^{\prime}}\equiv\nabla_{j^{\prime}}\sigma_{i^{\prime}}$.
A key quantity in our derivation is the derivative with respect to different points: $\sigma_{i^{\prime}j}\equiv\partial_{i^{\prime}}\sigma_{j}=\frac{\partial\sigma}{\partial x^{j}\partial x^{i^{\prime}}}$
and $\sigma_{ij^{\prime}}=\sigma_{j'i}$.The notation for coincidence limits of an arbitrary function is defined as:

\begin{equation}
\left[f\left(x,x^{\prime}\right)\right]=\underset{x\rightarrow x^{\prime}}{\lim}f\left(x,x^{\prime}\right).
\end{equation}

\noindent It is easy to see that

\begin{equation}
\left[\sigma(x,x')\right]=\left[\sigma_{i}\right]=\left[\sigma_{i'}\right]=0
\end{equation}

\noindent Remarkably, the coincidence limits of the second derivatives lead to the metric:

\begin{equation}
\left[\sigma_{ij^{\prime}}\right]\equiv\underset{x\rightarrow x'}{\lim}\partial_{x^{i}}\partial_{x^{j^{\prime}}}\left[\frac{1}{2}L_{\gamma_{A}}^{2}\left(x,x^{\prime}\right)\right]=-g_{ij}=-\left[\sigma_{ij}\right]=-\left[\sigma_{i'j^{\prime}}\right].\label{eq:derive metric}
\end{equation}

\noindent To better understand this formula, note that along a geodesic, the norm of the tangent vector $g_{ij}\dot{X}^{i}\dot{X}^{j}$ is constant. Thus, as $t\to0$, we find:

\begin{equation}
\sigma(x,x')=\frac{1}{2}L^{2}(x,x')=\frac{1}{2}\left[\int_{0}^{t}d\tau\:\sqrt{g_{ij}\dot{X}^{i}\dot{X}^{j}}\right]^{2}\approx\frac{1}{2}\,\lim_{t\to0}\,g_{ij}\frac{\Delta x^{i}}{t}\frac{\Delta x^{j}}{t}t^{2}\approx\frac{1}{2}g_{ij}\Delta x^{i}\Delta x^{j}.
\end{equation}

\noindent The advantage of $\sigma_{ij^{\prime}}$ over $\sigma_{ij}$ in eqn. (\ref{eq:derive metric})
is that we do not need to know the connection (geometry). Therefore, what we need to do is to determine the geodesic length of the \emph{yet-to-be-determined} dual geometry.
 
\begin{figure}[H]
\begin{centering}
\includegraphics[scale=0.5]{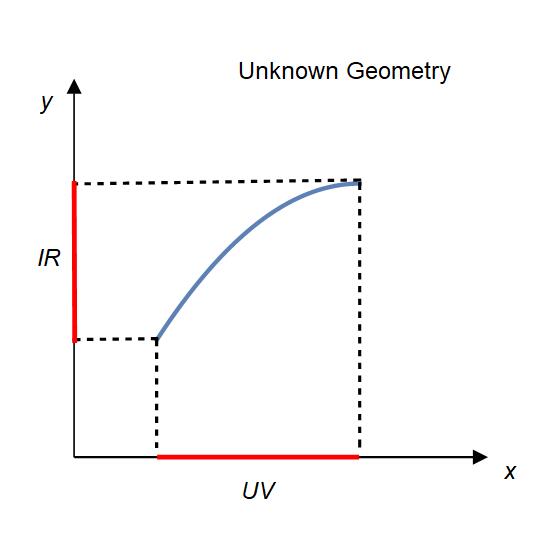}
\par\end{centering}
\centering{}\caption{\label{fig:UV_IR_fix}Two types of entanglement entropies are used to determine the leading behavior of the bulk geodesic.}
\end{figure}

It is important to note that the geodesic length here refers to a geodesic in the bulk, with its endpoints not fixed on the boundary. To address this, we need to use two types of known entanglement entropies to   fix the leading behavior of this geodesic length, as shown in Fig. (\ref{fig:UV_IR_fix}).

\vspace{2em}

\subsection*{1st Step: Two types of boundary-attached geodesics}
We begin by assuming no prior knowledge of the emerged geometry. From RT formula, our analysis relies only on the CFT EE, which we identify with the geodesic length, along with the holographic principle. In this work, we focus on a CFT$_2$ with infinite length. For simplicity, we first consider the static scenario, as generalizing the results to the covariant case is straightforward.
To begin, we define a quantity with length dimension:

\begin{equation}
R\equiv\frac{2G_{N}^{\left(3\right)}c}{3},
\end{equation}
where $c$ is the central charge of the CFT$_{2}$ and $G_{N}^{(3)}$
is the Newton constant in three dimensions. The EE of the $\mathrm{CFT_{2}}$
in the UV region is given by

\begin{equation}
S_{EE}^{UV}=\frac{c}{6}\log\left(\frac{\ell^{2}}{a^{2}}\right),\label{eq:UV EE}
\end{equation}

\noindent where $\ell=x-x^{\prime}$ represents the length of the entanglement sub-region in the CFT, and  $a$ is the UV cutoff or lattice spacing. It is important to note that this expression for the UV EE is valid only when
$\ell\gg a\rightarrow0$. Thus, the geodesic length of the dual geometry ending on the boundary is

\begin{equation}
L_{\gamma_{A}}^{UV}=4G_{N}^{\left(3\right)}S_{EE}=\left(\frac{2G_{N}^{\left(3\right)}c}{3}\right)\log\left(\frac{\ell^{2}}{a^{2}}\right)=R\log\left(\frac{\ell^{2}}{a^{2}}\right).\label{eq:UV geodesic}
\end{equation}

On the other hand, when the CFT$_{2}$ is perturbed by a relevant perturbation, the correlation length (IR cutoff) $\xi$ becomes finite. In this case, the UV EE expression (\ref{eq:UV EE}) is no longer valid for $\ell\ge\xi$. In the large $\ell$ IR region, with the condition $a\ll\xi\ll\ell$, the IR EE becomes independent of $\ell$, and is instead completely determined by the ratio of the IR and UV cutoffs \cite{Vidal:2002rm, Calabrese:2004eu},

\begin{equation}
S_{EE}^{IR}=\frac{c}{6}\log\left(\frac{\xi}{a}\right),\label{eq:IR EE}
\end{equation}

\noindent where $\xi\equiv1/m$ and $m$ is the mass gap of the perturbed
CFT$_{2}$. Note the crucial factor $1/6$, which plays a key role in deriving the expected geometry, a point we will return to later. This confirms that the RT formula also applies in the IR region \cite{Ryu:2006bv, Ryu:2006ef}.
Therefore, the geodesic length for the IR region, based on the perturbed CFT IR EE, is:
\begin{equation}
L_{\gamma_{A}}^{IR}=R\log\left(\frac{\xi}{a}\right).\label{eq:IR geo}
\end{equation}

\vspace{2em}

\subsection*{2nd Step: General expression for bulk geodesics}

It is widely recognized that the energy scale of a CFT gives rise to a hidden holographic dimension, denoted as $y$ in this paper \cite{Susskind:1998dq}. In the UV EE (\ref{eq:UV geodesic}),
$\ell$ represents the boundary dimension $x$, while $a$ introduces the holographic dimension  $y=a$. Eqn. (\ref{eq:UV geodesic}) is valid for geodesics that end on the boundary. Our goal is to generalize the expression $\frac{(x-x')^{2}}{y^{2}}$ in
(\ref{eq:UV geodesic}) to the bulk.  To generalize, we replace  $a^{2}$ with a regular function $h\left(x,x^{\prime},y,y^{\prime}\right)$.
The general extension of the proper length $\ell^{2}=(x-x')^{2}$ to include the holographic dimension $y$ is:

\begin{equation}
(x-x')^{2}\to\left(x-x^{\prime}\right)^{2}k\left(x,x^{\prime},y,y^{\prime}\right)+\left(y-y^{\prime}\right)^{2}p\left(x,x^{\prime},y,y^{\prime}\right),\label{eq:ell extension}
\end{equation}
where $k\left(x,x^{\prime},y,y^{\prime}\right)$,
$p\left(x,x^{\prime},y,y^{\prime}\right)$ are arbitrary regular functions. 
Note odd powers of $(x-x')$ or $(y-y')$ are prevented by the symmetry between $x(y)$ and $x'(y')$. Possible higher order even powers  of $(x-x')$ or $(y-y')$ are grouped into  $k\left(x,x^{\prime},y,y^{\prime}\right)$ and $p\left(x,x^{\prime},y,y^{\prime}\right)$. Cross terms  between $(x-x')$ and $(y-y')$ are excluded by the widely accepted viewpoint that the energy-generating holographic direction $y$ is perpendicular to the boundary direction $x$.

So, to match (\ref{eq:UV geodesic}) and (\ref{eq:IR geo}),   the geodesic can be generalized as
\begin{equation}
L_{\gamma_A}=R\log(\eta^2).\label{eq: boundary geo}
\end{equation}
%
%
%

\noindent for large $\eta^2$, where

\begin{equation}
\eta^{2}\equiv\frac{\left(x-x^{\prime}\right)^{2}k\left(x,x^{\prime},y,y^{\prime}\right)+\left(y-y^{\prime}\right)^{2}p\left(x,x^{\prime},y,y^{\prime}\right)}{h\left(x,x^{\prime},y,y^{\prime}\right)}.\label{eq:eta}
\end{equation}

\noindent Note that the functions $k$, $p$ and $h$ must be invariant under the exchange $\left(x^{\prime},y^{\prime}\right)\leftrightarrow\left(x,y\right)$. To recover eqn. (\ref{eq:UV geodesic}) in the limit $y=y'=a\to0$,
we must have: $k\left(x,x^{\prime},a,a\right)/h\left(x,x^{\prime},a,a\right)\sim1/a^{2}$
and $\left(y-y^{\prime}\right)^{2}p\left(x,x,y,y^{\prime}\right)\to0$.
These conditions ensure the generality of $\eta^{2}$. The expression for $\eta$ can be simplified by dividing $p(x,x',y,y')$
from both the numerator and denominator\footnote{It is equally good to cancel $k(x,x',y,y')$. It turns out the analysis
is similiar, though to some extent trickier.}:

\begin{equation}
\eta=\sqrt{\frac{f\left(x,x^{\prime},y,y^{\prime}\right)}{g\left(x,x^{\prime},y,y^{\prime}\right)}\left(x-x^{\prime}\right)^{2}+\frac{1}{g\left(x,x^{\prime},y,y^{\prime}\right)}\left(y-y^{\prime}\right)^{2}}.\label{eq:simple eta}
\end{equation}

\noindent On the other hand, when the endpoints of the geodesic approach
each other ($\eta\rightarrow0$), the geodesic length must vanish.
Therefore, the function $\chi\left(\eta\right)$ must take the form:

\begin{equation}
\chi\left(\eta\right)=1+C_{1}\eta^{\alpha}+\mathcal{O}\left(\eta^{2\alpha}\right),\qquad\eta\rightarrow0.
\end{equation}

\noindent In summary, we have:

\begin{equation}
L_{\gamma_{A}}=R\log\chi\left(\eta\right)=\begin{cases}
R\log\left(\eta^{2}+\mathcal{O}\left(\frac{1}{\eta^{2}}\right)\right), & \eta\to\infty,\\
R\log\left(1+C_{1}\eta^{\alpha}+\mathcal{O}\left(\eta^{2\alpha}\right)\right), & \eta\to0.
\end{cases}\label{eq:geo form}
\end{equation}

\noindent Here, $C_{i}$'s and $\alpha$ are constants to be determined. Thus, the next step is to fix the functions $f(x,x',y,y')$, $g(x,x',y,y')$, and the constant $\alpha$.

\vspace{2em}

\subsection*{3rd Step: Fixing the functions}

The next step is to determine the functions $f(x,x',y,y')$ and $g(x,x',y,y')$ from eqn. (\ref{eq:simple eta}):
\begin{equation}
\eta=\sqrt{\frac{f\left(x,x^{\prime},y,y^{\prime}\right)}{g\left(x,x^{\prime},y,y^{\prime}\right)}\left(x-x^{\prime}\right)^{2}+\frac{1}{g\left(x,x^{\prime},y,y^{\prime}\right)}\left(y-y^{\prime}\right)^{2}}.
\end{equation}

\noindent We will substitute this expression into the general form of the geodesic length, given by eqn. (\ref{eq:geo form}), to ensure it is consistent with the eqn. (\ref{eq:UV geodesic}) and eqn. (\ref{eq:IR geo}) in both the UV and IR regions, corresponding to boundary-attached geodesics.

\vspace{3em}

\noindent \textbf{A) $y=a$, $y^{\prime}=\xi$, and $x=x^{\prime}$ (Fixing $g(x,x',y,y')$)}

\vspace{1em}

\begin{figure}[H]
\begin{centering}
\includegraphics[scale=0.2]{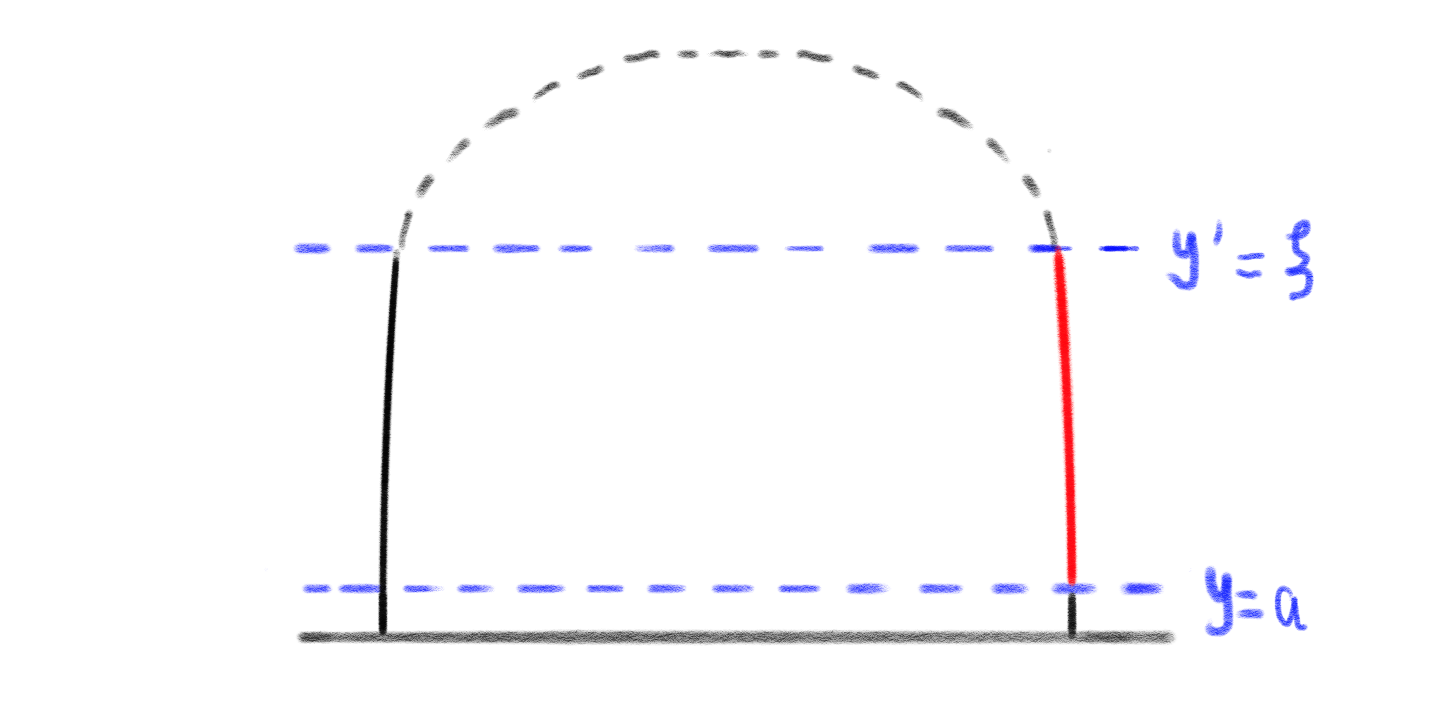}
\par\end{centering}
\centering{}\caption{\label{fig:IR_EE}The entanglement entropy with
IR cut-off gives the geodesic connecting $y=a$, $y^{\prime}=\xi$
and $\Delta x=0$.}
\end{figure}

\noindent In this case, the IR EE condition is given by eqn. (\ref{eq:IR geo}):
\begin{equation}
L_{\gamma_{A}}^{IR}=R\log\left(\frac{\xi}{a}\right).
\end{equation}

\noindent Here, both $a$ and $\xi$ are energy scales corresponding to the two points in the holographic direction, $y=a$ and $y^{\prime}=\xi$, respectively. Additionally, $x-x^{\prime}\rightarrow0$.
The geometry corresponding to this scenario is illustrated in  Fig. (\ref{fig:IR_EE}).
Referring to the large $\eta$ limit in the general expression (\ref{eq:geo form}) and (\ref{eq:simple eta}), we find:

\begin{equation}
L_{\gamma_{A}}=R\log\left(\frac{\left(y-y^{\prime}\right)^{2}}{g\left(x,x^{\prime},y,y^{\prime}\right)}\right)\simeq R\log\left(\frac{\left(a-\xi\right)^{2}}{g\left(x,x,a,\xi\right)}\right),
\end{equation}

\noindent where we take the limit $\triangle x\rightarrow0$ and use
$y\left(=a\right)\ll y^{\prime}\left(=\xi\right)$. In this limit,
for the numerator, we have:

\begin{equation}
\left(y-y^{\prime}\right)^{2}=\left(a-\xi\right)^{2}\sim\xi^{2}.
\end{equation}

\noindent Comparing this with

\begin{equation}
L_{\gamma_{A}}^{IR}=R\log\left(\frac{\xi}{a}\right),
\end{equation}

\noindent we can now fix the leading behavior of the denominator:

\begin{equation}
g\left(x,x,a,\xi\right)=a\xi\left(1+\mathcal{O}\left(\frac{\triangle x}{\xi}\right)^{2}\right),\label{eq:g limit}
\end{equation}

\noindent where $\xi$ is introduced to make the expansion parameter
dimensionless. Note that $\triangle x\rightarrow0$ and $\xi\rightarrow\infty$
in the expansion. The function with arbitrary endpoints can be extracted
from this limit result (\ref{eq:g limit}):
\noindent \begin{center}
\noindent\fbox{\begin{minipage}[t]{1\columnwidth \fboxsep \fboxrule}%
\noindent \begin{center}
\textbf{$\triangle x$ is finite and $\xi\rightarrow\infty$:}
\par\end{center}
\begin{equation}
g\left(x,x^{\prime},y,y^{\prime}\right)=yy^{\prime}\left(1+\left(\frac{\triangle x}{\xi}\right)^{2}\right),\label{eq:g full}
\end{equation}
\end{minipage}}
\par\end{center}

\vspace{3em}

\noindent \textbf{B) $y=y^{\prime}=a\ll\ell\ll\xi$ (Fixing $f(x,x',y,y')$)}

\vspace{1em}

\begin{figure}[H]
\begin{centering}
\includegraphics[scale=0.3]{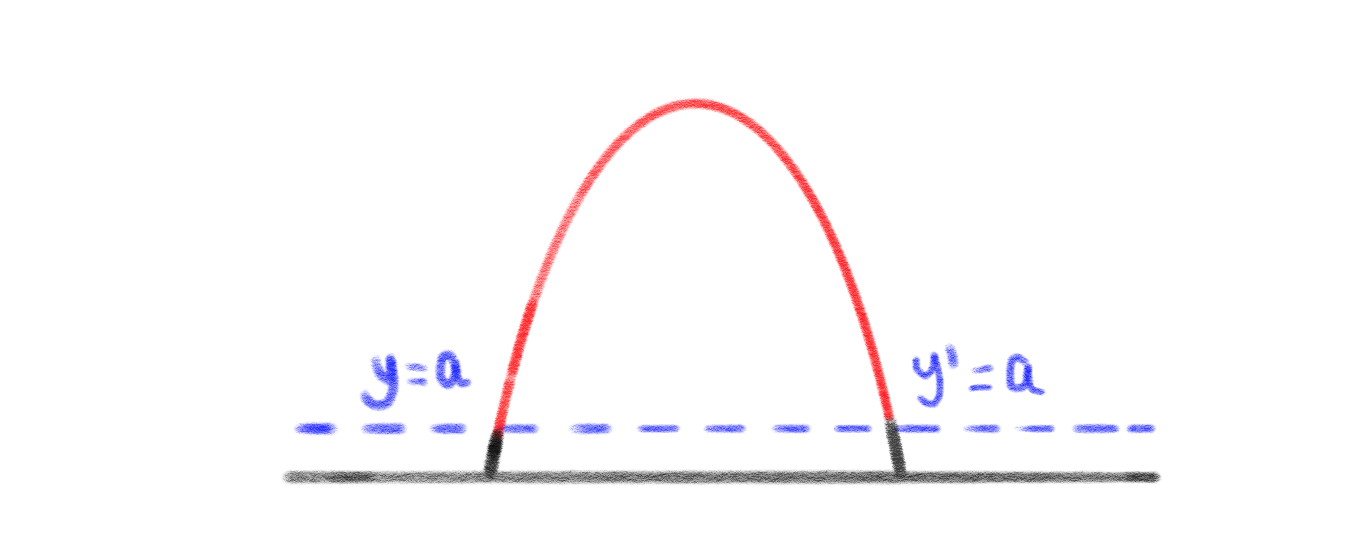}
\par\end{centering}
\centering{}\caption{\label{fig:UV_EE}The entanglement entropy with UV cut-off supports the geodesic connecting $y=y^{\prime}=a$ with arbitrary $\Delta x$.}
\end{figure}

\noindent In this scenario, we are in the UV regime, and the EE is given by: 
\begin{equation}
L_{\gamma_{A}}^{UV}=R\log\left(\frac{\ell^{2}}{a^{2}}\right).\label{eq:UV geodesic 2}
\end{equation}

\noindent Since $y=y^{\prime}=a$, we have $y-y^{\prime}=0$, and the general geodesic length expression becomes:

\begin{equation}
L_{\gamma_{A}}=R\log\left(\frac{f\left(x,x^{\prime},y,y^{\prime}\right)}{g\left(x,x^{\prime},y,y^{\prime}\right)}\left(x-x^{\prime}\right)^{2}\right)\simeq R\log\left(\frac{f\left(x,x^{\prime},a,a\right)}{g\left(x,x^{\prime},a,a\right)}\left(x-x^{\prime}\right)^{2}\right).
\end{equation}

\noindent
For consistency with eqn. (\ref{eq:UV geodesic 2}), we must have: $f\left(x,x^{\prime},a,a\right)/g\left(x,x^{\prime},a,a\right)\sim1/a^{2}$. This condition ensures that we recover the expected UV behavior as $y=y'=a\to0$.
The corresponding geometry is shown in 
Fig. (\ref{fig:UV_EE}). From the determined function $g(x,x',y,y')$ in eqn. (\ref{eq:g full}), we immediately obtain

\begin{equation}
f\left(x,x^{\prime},0,0\right)=1+\mathcal{O}\left(\frac{\triangle x}{\xi}\right)^{2}.
\end{equation}

\noindent Here, $\triangle x$ is finite, and $y=y^{\prime}=a\ll \ell \ll\xi$.
In this limit, we can expand the function $f\left(x,x^{\prime},y,y^{\prime}\right)$
as
\noindent \begin{center}
\noindent\fbox{\begin{minipage}[t]{1\columnwidth\fboxsep\fboxrule}%
\noindent \begin{center}
\textbf{$\triangle x$ is finite and }$y$, $y^{\prime}\ll\xi\rightarrow\infty$\textbf{:}
\par\end{center}
\begin{equation}
f\left(x,x^{\prime},y,y^{\prime}\right)=\left(1+\mathcal{O}\left(\frac{\triangle x}{\xi}\right)^{2}\right)+\mu_{1}\left(x,x^{\prime}\right)\left(\frac{y}{\xi}+\frac{y^{\prime}}{\xi}\right)+\mu_{2}\left(x,x^{\prime}\right)\left(\frac{y}{\xi}+\frac{y^{\prime}}{\xi}\right)^{2}+\mathcal{O}\left(\frac{y}{\xi},\frac{y^{\prime}}{\xi}\right)^{3},\label{eq:f full}
\end{equation}
\end{minipage}}
\par\end{center}

\noindent where $\mu_{i}\left(x,x^{\prime}\right)$ are regular functions.
Now, recall the expression for the geodesic length:

\begin{equation}
L_{\gamma_{A}}=R\log\left(1+C_{1}\eta^{\alpha}+\mathcal{O}\left(\eta^{2\alpha}\right)\right),\qquad\eta\rightarrow0,
\end{equation}

\noindent where

\begin{equation}
\eta=\sqrt{\frac{\left(x-x^{\prime}\right)^{2}f\left(x,x^{\prime},y,y^{\prime}\right)+\left(y-y^{\prime}\right)^{2}}{g\left(x,x^{\prime},y,y^{\prime}\right)}}.
\end{equation}

\noindent As $\eta\rightarrow0$ ($\triangle x\rightarrow0$ and $\triangle y\rightarrow0$)
and $\xi\rightarrow\infty$, if we require $y$, $y^{\prime}\ll\xi\rightarrow\infty$,
the expressions for $g\left(x,x^{\prime},y,y^{\prime}\right)$ (\ref{eq:g full})
and $f\left(x,x^{\prime},y,y^{\prime}\right)$ (\ref{eq:f full})
still hold. Therefore, we have
\noindent \begin{center}
\noindent\fbox{\begin{minipage}[t]{1\columnwidth\fboxsep\fboxrule}%
\noindent \begin{center}
\textbf{$\triangle x\rightarrow0$, }$\triangle y\rightarrow0$\textbf{
and }$y$, $y^{\prime}\ll\xi\rightarrow\infty$\textbf{:}
\par\end{center}
\begin{eqnarray}
L_{\gamma_{A}} & = & R\log\left(1+C_{1}\left(\frac{\left(x-x^{\prime}\right)^{2}\left(\left(1+\mathcal{O}\left(\frac{\triangle x}{\xi}\right)^{2}\right)+\mu_{1}\left(x,x^{\prime}\right)\left(\frac{y}{\xi}+\frac{y^{\prime}}{\xi}\right)+\ldots\right)+\left(y-y^{\prime}\right)^{2}}{yy^{\prime}\left(1+\mathcal{O}\left(\frac{\triangle x}{\xi}\right)^{2}\right)}\right)^{\alpha/2}+\ldots\right)\nonumber \\
 & = & R\log\left(1+C_{1}\left(\frac{\left(x-x^{\prime}\right)^{2}+\left(y-y^{\prime}\right)^{2}}{yy^{\prime}}\right)^{\alpha/2}+\mathcal{O}\left[\left(\frac{\left(\triangle x\right)^{2}}{yy^{\prime}}\right)^{1/2},\left(\frac{\left(\triangle y\right)^{2}}{yy^{\prime}}\right)^{1/2},\frac{y}{\xi},\frac{y^{\prime}}{\xi},\frac{\triangle x}{\xi}\right]^{2}\right),
\end{eqnarray}
\end{minipage}}
\par\end{center}

\noindent where the order of $\mathcal{O}\left[A,B,\ldots\right]^{2}$
denotes the higher order of $\mathcal{O}\left[A^{2}\right]$, $\mathcal{O}\left[B^{2}\right]$,
$\mathcal{O}\left[AB\right]$....

\vspace{2em}

\subsection*{4th Step: Extracting the metric via Synge's world function}

We now aim to calculate the metric by using the geodesic length. To do this, we can apply the following relation based on Synge's world function:

\begin{equation}
\left[\sigma_{ij^{\prime}}\right]=\underset{x^{j^{\prime}}\rightarrow x^{i}}{\lim}\partial_{x^{i}}\partial_{x^{j^{\prime}}}\left[\frac{1}{2}L_{\gamma_{A}}^{2}\left(x,x^{\prime}\right)\right]=-g_{ij},
\end{equation}

\noindent where we need to determine the value of  $\alpha$. Through some computation, we obtain the following results:

\begin{equation}
g_{ij}=\begin{cases}
{\rm divergent}, & \alpha<1,\\
{\rm valid}, & \alpha=1,\\
0, & \alpha>1.
\end{cases}
\end{equation}

\noindent This implies that only  $\alpha=1$ yields a physical solution. Therefore, we can now extract the nonvanishing components of the metric:

\begin{eqnarray}
g_{xx} & = & -\frac{1}{2}\:\underset{\left(x^{\prime},y^{\prime}\right)\rightarrow\left(x,y\right)}{\lim}\:\partial_{x}\partial_{x^{\prime}}L_{\gamma_{A}}^{2}=C_{1}^{2}\left(1+\mu_{1} \left(\frac{y}{\xi}\right)+\ldots\right)\frac{R^{2}}{y^{2}},\nonumber \\
g_{yy} & = & -\frac{1}{2}\:\underset{\left(x^{\prime},y^{\prime}\right)\rightarrow\left(x,y\right)}{\lim}\:\partial_{y}\partial_{y^{\prime}}L_{\gamma_{A}}^{2}=C_{1}^{2}\frac{R^{2}}{y^{2}}.
\end{eqnarray}

\noindent The functions $s\left(x,x^{\prime},0,0\right)$
do not contribute to the metric. Thus, the background metric is:

\begin{equation}
ds^{2}=C_{1}^{2}\frac{R^{2}}{y^{2}}\left[\left(1+\mu_{1}\left(\frac{y}{\xi}\right)+\mu_{2}\left(\frac{y}{\xi}\right)^{2}+\ldots\right)dx^{2}+dy^{2}\right].
\end{equation}

\noindent This is the asymptotic form of static AdS$_{3}$ with
radius $C_{1}R$. This is the asymptotic form of static $\mathrm{AdS_{3}}$
and the central charge $c$ of $\mathrm{CFT_{2}}$ \cite{Brown:1986}:
$c=\frac{3R_{AdS}}{2G_{N}^{\left(3\right)}}$, we identify:

\begin{equation}
C_{1}=1,\qquad R=R_{AdS}.
\end{equation}

\noindent  Therefore, we finally determine the leading terms of the series (\ref{eq:geo form})\footnote{Remarkably, if we know the AdS geometry and calculate the
geodesic length, the length takes the following form:

\[
L_{\gamma_{A}}=R\log\left(  1+\frac{1}{2}\eta^2 + \frac{1}{2}\eta^2 \sqrt{1+4\eta^{-2}} \right),\qquad\eta^2 = \frac{\left(x-x^{\prime}\right)^{2}+\left(y-y^{\prime}\right)^{2}}{yy^{\prime}},
\]

This expression completely agrees with the leading behaviors
we derived in the two limits (\ref{eq:geo form-1}).
}:

\begin{equation}
L_{\gamma_{A}}=R_{AdS}\log\chi\left(\eta\right)=\begin{cases}
R\log\left(\eta^{2}+\mathcal{O}\left(\frac{1}{\eta^{2}}\right)\right), & \eta\to\infty,\\
R\log\left(1+\eta+\mathcal{O}\left(\eta^{2}\right)\right), & \eta\to0.
\end{cases}\label{eq:geo form-1}
\end{equation}

\noindent Thus, the metric becomes:

\begin{equation}
ds^{2}=\frac{R_{AdS}^{2}}{y^{2}}\left[\left(1+\mu_{1}\left(\frac{y}{\xi}\right)+\mu_{2}\left(\frac{y}{\xi}\right)^{2}+\ldots\right)dx^{2}+dy^{2}\right].
\end{equation}
To recover pure AdS$_3$, we introduce the IR cut-off scale defined as:

\begin{equation}
\xi=\frac{1}{m}.
\end{equation}

\noindent In the massless limit $m\rightarrow0$ (or equivalently 
$\xi\rightarrow\infty$), we recover the pure AdS$_3$ metric:

\begin{equation}
ds^{2}=\frac{R_{AdS}^{2}}{y^{2}}\left(dx^{2}+dy^{2}\right).
\end{equation}

When including the time-like direction, the UV EE is:

\begin{equation}
S_{EE}^{UV}=\frac{c}{6}\log\left(\frac{\ell^{2}-\left(\triangle t\right)^{2}}{a^{2}}\right).
\end{equation}
The procedure follows the same pattern as in the static case. To generalize, when writing $\eta$ as in eqn. (\ref{eq:simple eta}),
we introduce an arbitrary function multiplying $(\Delta t)^{2}$. After performing a similar analysis, we obtain the covariant AdS metric:

\begin{equation}
ds^{2}=\frac{R_{AdS}^{2}}{y^{2}}\left[-G(t,x,y)\,dt^{2}+F(t,x,y)\,dx^{2}+dy^{2}\right],\label{eq:covariant AdS}
\end{equation}

\noindent where

\begin{eqnarray*}
G(t,x,y) & = & 1+\rho_{1}\left(\frac{y}{\xi}\right)+\rho_{2}\left(\frac{y}{\xi}\right)^{2}+\ldots\\
F(t,x,y) & = & 1+\mu_{1}\left(\frac{y}{\xi}\right)+\mu_{2}\left(\frac{y}{\xi}\right)^{2}+\ldots
\end{eqnarray*}

\noindent As $\xi\rightarrow\infty$, we recover pure AdS$_3$:

\begin{equation}
ds^{2}=\frac{R_{AdS}^{2}}{y^{2}}\left(-dt^{2}+dx^{2}+dy^{2}\right).
\end{equation}

\noindent The result is consistent with the holographic RG flow
discussed in \cite{Liu:2012eea}. To illustrate this, we consider the action for a bulk scalar field $\phi_{a}$:

\begin{equation}
I=\frac{1}{2\kappa^{2}}\int d^{3}x\sqrt{-g}\left[R-\frac{1}{2}G_{ab}\partial\phi_{a}\partial\phi_{b}-V\left(\phi_{a}\right)\right],
\end{equation}

\noindent where $G_{ab}$ is the metric of the internal space of the scalar field. When the potential $V\left(\phi_{a}\right)$ reaches its critical value  $V\left(0\right)=-\frac{2}{R^{2}}$, the solution for the metric is:

\begin{equation}
ds^{2}=\frac{R_{AdS}^{2}}{y^{2}}\left[-dt^{2}+dx^{2}+F\left(y\right)dy^{2}\right].
\end{equation}

\noindent Near the boundary $y=0$, we obtain the asymptotic expansion:

\begin{equation}
\phi_{a}\left(y\right)\rightarrow0,\qquad F\left(y\right)=1+\mu^{2\alpha}y^{2\alpha}+\cdots,\qquad y\rightarrow0,
\end{equation}

\noindent where $\mu$ is some mass scale. This is a special case of eqn. (\ref{eq:covariant AdS}) under the condition $G(t,x,y)=F(t,x,y)$.

\section{A necessary condition for a holographic CFT$_{2}$}

We begin by recalling the two types of  entanglement entropies and their corresponding geodesic lengths in the context of AdS/CFT:

\begin{eqnarray}
S_{EE}^{UV}=\frac{c}{6}\log\left(\frac{\ell}{a}\right)^{2} & \rightarrow & L_{\gamma_{A}}^{UV}=R\log\left(\frac{\ell}{a}\right)^{2},\\
S_{EE}^{IR}=\frac{c}{6}\log\left(\frac{\xi}{a}\right) & \rightarrow & L_{\gamma_{A}}^{IR}=R\log\left(\frac{\xi}{a}\right).
\end{eqnarray}

\noindent These results follow directly from the AdS/CFT correspondence. However, if we do not assume the AdS/CFT framework, meaning we do not have a direct relation between the AdS radius and the central charge, it becomes difficult to explain why the first formula has a squared term while the second formula does not.
To address this, we propose that the IR EE is multiplied by a constant factor
$\mathcal{A}$, and identify this factor as modifying the corresponding geodesic length:

\begin{equation}
S_{EE}^{IR}=\mathcal{A}\:\frac{c}{6}\log\left(\frac{\xi}{a}\right)\Longrightarrow L_{\gamma_{A}}^{IR}=R\log\left(\frac{\xi}{a}\right)^{\mathcal{A}}.\label{eq:NEW IR EE}
\end{equation}
It is straightforward to verify that for this modified IR EE to be compatible with the UV EE in the form given by eqn. (\ref{eq:UV geodesic}), the function $\eta$ must take the form:

\begin{equation}
\eta=\sqrt{\frac{\left(x-x^{\prime}\right)^{2}f\left(x,x^{\prime},y,y^{\prime}\right)+\left(y-y^{\prime}\right)^{2\mathcal{A}}}{g\left(x,x^{\prime},y,y^{\prime}\right)}},\label{eq:new simple eta}
\end{equation}
where $g(x,x',y,y')=(yy')^{\mathcal{A}}(1+\Delta x^{2}s(x,x',y,y')/\xi^{2}+\cdots)$,$\:$ and
$f(x,x',0,0)=(yy')^{\mathcal{A}-1}(1+\Delta x^{2}s(x,x',0,0)/\xi^{2}+\cdots)$.
When substituting this expression for $\eta$ into the eqn. (\ref{eq:derive metric}) for the metric, we find that
$g_{yy}$ behaves as follows:

\begin{equation}
g_{yy}=\begin{cases}
{\rm 0}, & \mathcal{A}=0,1/2;\:\mathcal{A}>1,\\
{\rm valid}, & \mathcal{A}=1,\\
{\rm divergence}, & {\rm otherwise}.
\end{cases}
\end{equation}
Thus, we conclude that there is no dual gravity for $\mathcal{A}\not=1$,
which provides a necessary condition for determining whether a CFT$_{2}$ is holographic. Furthermore, $\mathcal{A}=1$ is a necessary condition for any holographic CFT. This is consistent with the fact that, for a single interval, although there are two boundary points, the holographic dual IR EE only counts one point.

\vspace{5mm}

\noindent {\bf Acknowledgements}
We are deeply indebt to Bo Ning for many illuminating discussions and suggestions. This work cannot be done without her help. We are also very grateful to S. Kim, J. Lu,  H. Nakajima and S. He for very helpful discussions and suggestions. This work is supported in part by the NSFC (Grant No. 12105191, 12275183 and 12275184).


\end{document}